# Goos-Hänchen shift at the reflection of light from the complex structures composed of superconducting and dielectric layers

Yu. S. Dadoenkova,[1,2,3a)] N. N. Dadoenkova,[1,3] I. L. Lyubchanskii,[3] and Y.P. Lee[4]

[1]*Ulyanovsk State University, 432017, Ulyanovsk, Russian Federation*

[2]*Institute of Electronic and Information Systems, Novgorod State University, 173003, Veliky Novgorod, Russian Federation*

[3]*Donetsk Physical and Technical Institute of the National Academy of Science of Ukraine, 83114, Donetsk, Ukraine*

[4]*Quantum Photonic Science Research Center (q-Psi) and Hanyang University, 04763, Seoul, South Korea*

The Goos-Hänchen effect of light reflected from sandwich (three-layered) structures composed of a superconducting $YBa_2Cu_3O_7$ film and two different dielectric films is investigated theoretically. It has been shown that optical anisotropy of $YBa_2Cu_3O_7$ film, as well as its positions in the three-layer specimen, strongly effects on the lateral shift values. We have shown that, for all positions of the superconducting film in the three-layered structure, variation of temperature makes possible to control the values of the lateral shift of TE-polarized light at the incidence angles close to pseudo-Brewster angles, whereas for TM-polarized light the lateral shift is only significant at grazing incidence.

## I. INTRODUCTION

The effect of lateral shift of light at the reflection from the interface of materials with different refractive indices, also known as Goos-Hänchen effect, is widely studied both theoretically and experimentally started from the famous papers by F. Goos and H. Hänchen [1, 2]. Until now investigation of the Goos-Hänchen and related Imbert-Fedorov effects in different media is a topical problem, as shown in recent review papers [3, 4]. Phenomenon of the lateral shift at the reflection and transmission of light was studied for electro-optic [5 – 7] and magneto-optic materials [8 – 10], for photonic crystals [11 – 17], plasmonic structures [18 – 20], graphene [21 – 23], as well as for metamaterials [24 – 26]. In recent publications the successful applications of the Goos-Hänchen shift measurements to detect E. Coli O157: H7 concentration [27] and chemical vapors [28] have been reported. It is interesting to investigate the effect of the lateral shift for multilayered structure containing superconducting (SC) films because nowadays a very intensive research in the area of SC metamaterials and SC photonic crystals is rapidly developing as reflected in recent review papers [29, 30].

_________________________

a) Electronic mail: yulidad@gmail.com

It has been shown theoretically and confirmed experimentally [31] that the Goos-Hänchen shift in metal for TE-polarized light is negative and much larger by modulus, than the positive shift for TM-polarized light. On the other hand, the Goos-Hänchen shift at light reflection from a dielectric film backed by a metal could be both positive and negative depending on the incidence angle and material parameters of the system [32]. Negative Goos-Hänchen shift was also reported for reflected beam emerged from a grounded slab with both negative permittivity and permeability [33]. We anticipate that the lateral shift of light reflected from a high-temperature ceramic SC demonstrates different behavior for TE- and TM-polarized waves, influenced by anisotropy of optical properties and temperature dependence of the SC medium.

In this paper, we theoretically investigate the effect of the lateral shift for three-layer complex structure composed of SC and two dielectric films, which are characterized by different refractive indexes, *i.e.* asymmetric sandwich structure. We calculate and analyze dependence of the Goos-Hänchen shift on the SC layer thickness and mutual location of the SC and dielectric layers; as well we examine the possibility of governing the lateral shift by tuning the incidence angle, temperature and incident light frequency.

## II. MODEL AND GENERAL DESCRIPTION

Let us consider the reflection of an electromagnetic wave from the sandwich structure consisting of a SC film $YBa_2Cu_3O_7$ (YBCO) and two dielectric films of strontium titanate $SrTiO_3$ (STO) and aluminum oxide $Al_2O_3$ (AlO) with thicknesses $d_{YBCO}$, $d_{STO}$, and $d_{AlO}$, respectively, as depicted in Fig. 1. We choose these dielectric materials because they are widely used as substrates for the SC films. The layers of the structure are located parallel to *xy*-plane, and the *z*-axis is perpendicular to the interfaces. We investigate and compare the lateral shift of light reflected from all possible three-layer combinations of these materials: YBCO/STO/AlO, YBCO/AlO/STO [Fig. 1(a)], STO/YBCO/AlO, AlO/YBCO/STO [Fig. 1(b)], and STO/AlO/YBCO, AlO/STO/YBCO [Fig. 1(c)].

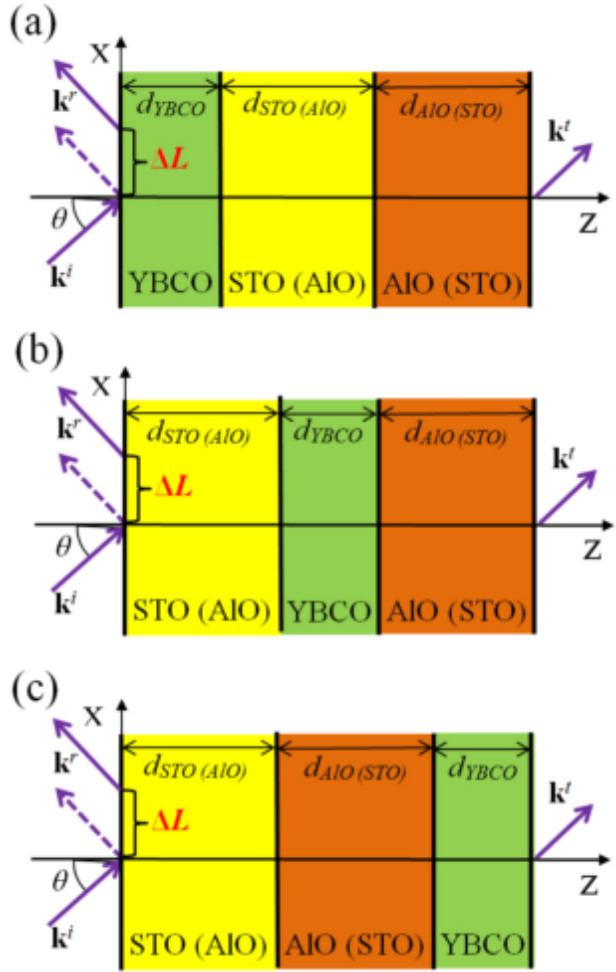

FIG. 1. Schematic of the Goos-Hänchen effect at light reflection from a sandwich superconductor/dielectric structures: (a) YBCO/STO/AlO (YBCO/AlO/STO); (b) STO/YBCO/AlO (AlO/YBCO/STO); (c) STO/AlO/YBCO (AlO/STO/YBCO). Here $\mathbf{k}^i$, $\mathbf{k}^r$ and $\mathbf{k}^t$ denote wavevectors of the incident, reflected and transmitted waves, respectively, and $\theta$ is the incidence angle.

The electromagnetic wave of angular frequency $\omega$ is running from the vacuum under the incidence angle $\theta$ (see Fig. 1). The reflected beam undergoes a lateral shift $\Delta L$ from the position following out from the geometrical optics (see solid and dashed arrows in Fig. 1). The wavevectors of the incident, reflected and transmitted waves are denoted as $\mathbf{k}^i$, $\mathbf{k}^r$ and $\mathbf{k}^t$, respectively. We use the stationary phase method to study the Goos-Hänchen effect. Let us assume that the incident electromagnetic wave can be represented as a wavepacket of a Gaussian shape, with a characteristic length $\Delta k_x / k_x \ll 1$, then the reflected beam will show the space shift $\Delta L$ relatively to the incident wavepacket's position

$$\Delta L = -\frac{\partial \psi}{\partial k_x}, \quad \psi = \arctan \frac{\text{Im}(R)}{\text{Re}(R)}, \qquad (1)$$

where $\psi$ is a phase difference between the reflected and incident waves, Im($R$) and Re($R$) are imaginary and real parts of the reflection coefficient $R$, $k_x$ is the $x$-component of the wavevector.

For the SC layer, we will distinguish two cases of the crystallographic axes orientation relative to the incidence plane. When YBCO film is epitaxially grown on the substrate so that its crystallographic *a*-axis and *c*-axis coincide with the *x*- and *y*-axes, respectively, the frequency- and temperature-dependent dielectric permittivity tensor has nonzero diagonal components $\varepsilon_{xx}(\omega,T) = \varepsilon_{zz}(\omega,T)$ and $\varepsilon_{yy}(\omega,T)$. It should be noted that both real and imaginary parts of $\varepsilon_{xx}(\omega,T)$ are about two orders of magnitude larger than the corresponding ones of $\varepsilon_{yy}(\omega,T)$, so the SC film possesses strong anisotropy of optical properties in *xy*-plane [34]. In this case, the electromagnetic radiation in the SC film can be presented by independent TE- and TM-modes with the *z*-components of the wavevectors

$$k_z^{TE} = \sqrt{k_0^2 \varepsilon_{yy} - k_x^2}, \qquad k_z^{TM} = \sqrt{k_0^2 \varepsilon_{xx} - k_x^2}, \qquad (2)$$

respectively, $k_0$ is the wave number of the light in vacuum.

When *a*- and *c*-axes of YBa$_2$Cu$_3$O$_7$ coincide with the *x*- and *z*-axes, respectively, it leads to isotropy of the dielectric permittivity in *xy*-plane, and the *z*-components of the wavevectors of TE- and TM-modes are

$$k_z^{TE} = \sqrt{k_0^2 \varepsilon_{xx} - k_x^2}, \qquad k_z^{TM} = \sqrt{k_0^2 \varepsilon_{xx} - k_x^2 \varepsilon_{xx}/\varepsilon_{yy}}. \qquad (3)$$

The frequency-dependent components of the dielectric permittivity tensor of the SC layer are calculated according to Ref. [34]. It should be noted that in the optical and near infrared regimes the electrodynamic properties of YBCO can be described by the dielectric permittivity tensor only, *i.e.* the magnetic permeability is assumed to be $\mu = 1$. We expect that the anisotropy of the permittivity of YBCO results in different behavior of the Goos-Hänchen shift of TE- and TM-polarized waves.

### III. RESULTS OF THE NUMERICAL CALCULATIONS

We calculated the lateral shift at the Goos-Hänchen effect for the TE- and TM-polarized incident light for the SC layer of thickness $d_{YBCO}$ up to 70 nm. This limit in the thickness of SC film is due to the strong absorption in thick SC layers as was shown in Ref. [34]. The thicknesses of the STO and AlO layers are taken to be $d_{STO} = 4$ μm and $d_{AlO} = 6$ μm because these materials are transparent in the visible regime, and their refractive indices are $n_{STO} = 2.437$ and $n_{AlO} = 1.767$, respectively [35]. The results of the numerical calculations are presented in Figs. 2 – 6.

## A. Goos-Hänchen shift as function of the incidence angle for different mutual locations of SC and dielectric layers

First, we consider the Goos-Hänchen effect for the sandwich systems with different positions of the SC layer: the SC film on the top of STO/AlO-bilayer [structures YBCO/STO/AlO and YBCO/AlO/STO depicted in Fig. 1(a)]; the SC film between $SrTiO_3$ and $Al_2O_3$ films [STO/YBCO/AlO and AlO/YBCO/STO shown in Fig. 1(b)]; the SC slab on the bottom of $SrTiO_3$/$Al_2O_3$-bilayer [STO/ALO/YBCO and AlO/STO/YBCO, see Fig. 1(c)]. We focus on the case of anisotropic permittivity of the SC film in *xy*-plane.

In Fig. 2 the normalized lateral shifts $\Delta X_{TE} = \Delta L/\lambda$ (where $\lambda$ is the wavelength of the incident electromagnetic wave) of TE-polarized incident light are given as functions of the incidence angle $\theta$ for the fixed angular frequency $\omega=100$ rad×THz and temperature $T = 5$ K. In the structures where the electromagnetic wave first impinges the $SrTiO_3$ layer and then comes to the $Al_2O_3$ layer (such systems as YBCO/STO/AlO, STO/YBCO/AlO, and STO/AlO/YBCO) the lateral shifts $\Delta X_{TE}$ exhibit maxima at some incidence angles ($\theta \approx 43°$), as one can see from Fig. 2(a). The maximal values of $\Delta X_{TE}$ can reach several tens of wavelengths. It should be noted that these maxima correspond to the reflection under the pseudo-Brewster angles, when the reflection coefficient of TE-polarized light is minimal but nonzero, and its phase experiences a distinct fast variation with $k_x$, which result in a sharp extremum of the lateral shift.

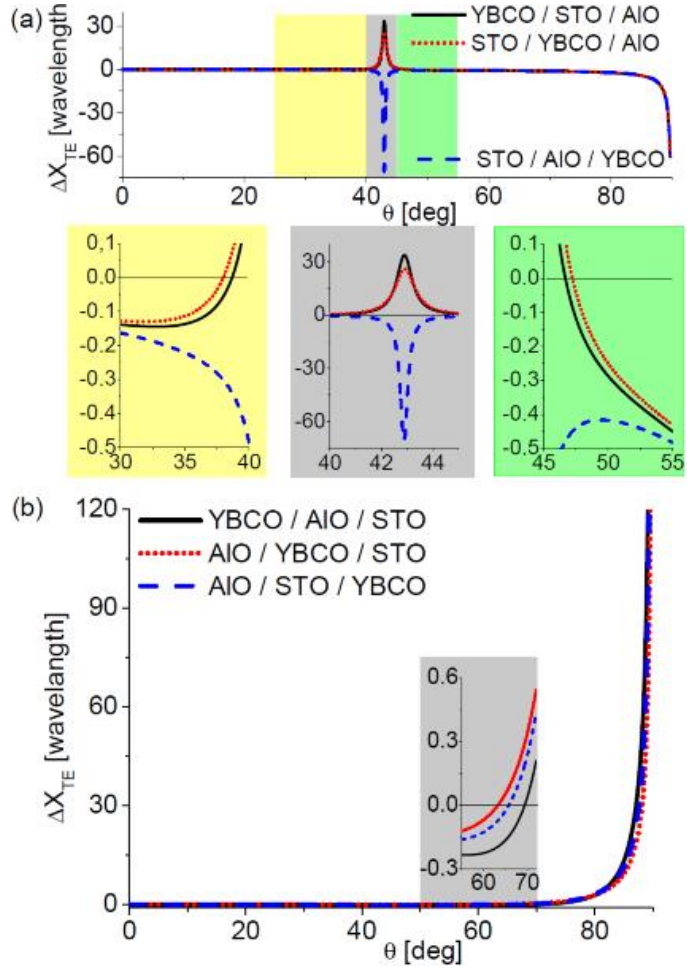

FIG. 2. Dimensionless lateral shifts $\Delta X_{TE}$ vs incidence angle $\theta$ for different positions of the SC film for TE-polarized electromagnetic wave incident in (a) STO, AlO and (b) AlO, STO dielectric layers (in order of light passing though the structure). The thickness of the SC film is $d_{YBCO} = 50$ nm, the angular frequency $\omega = 100$ rad×THz.

For the structure STO/AlO/YBCO the Goos-Hänchen shift is negative for all incidence angle range [dashed line in Fig. 2(a)], whereas for the systems YBCO/STO/AlO and STO/YBCO/AlO the lateral shift peaks around the pseudo-Brewster angles are positive (solid and dotted curves in Fig. 2(a), respectively), and for $\theta$ far from the pseudo-Brewster angles $\Delta X_{TE}$ (at the grazing incidence) are negative and their values are about several tenth of wavelength. It has been shown in Ref. [32] that at light reflection from a dielectric film backed by a metal, a negative lateral shift is due to the interference of the reflected waves from the dielectric-air and dielectric-metal interfaces. In our sandwich systems the direction of the lateral shifts is governed by the interference of the waves reflected from each interface separating dielectric films and YBCO layer. As one can see from the left and right insets in Fig. 2(a), $\Delta X_{TE}$ for the aforementioned two structures reaches zero at some $\theta$ in the vicinity of the pseudo-Brewster angles ($\theta \approx 38°$ and $\theta \approx 46°$ for YBCO/STO/AlO and $\theta \approx 37°$ and $\theta \approx 48°$ for STO/YBCO/AlO). For the grazing incidence, $\Delta X_{TE}$ is negative and practically identical for all structures of this type.

For the structures where the light first impinges the AlO layer and then comes to the STO film (namely, YBCO/AlO/STO, AlO/YBCO/STO, and AlO/STO/YBCO), $\Delta X_{TE}$ demonstrates no peaks as for these structures there are no pseudo-Brewster angles. The lateral shifts' behavior is similar to the case of STO → AlO propagation only at grazing incidence [see Fig. 2(b)]. For the incidence angles far from the grazing incidence, $\Delta X_{TE}$ are negative and their values are about several tenth of wavelength. At some $\theta$ around 63°, 65°, and 68° for AlO/YBCO/STO, AlO/STO/YBCO, and YBCO/AlO/STO, respectively [dotted, dashed, and solid lines in Fig. 2(b)], $\Delta X_{TE}$ becomes zero, and at grazing incidence the lateral shifts grow intensely with further $\theta$ increase to the values of some tens wavelengths.

The analogous dependencies of the lateral shift $\Delta X_{TM}$ for TM-polarized light are presented in Fig. 3. The lateral shift for all considered structures becomes significant only at the grazing incidence. For these configurations the pseudo-Brewster condition is only satisfied for STO/YBCO/AlO system at $\theta \approx 85°$, which results in a positive maximum of the lateral shift, as shown in Fig. 3(a) with dotted line. The values of $\Delta X_{TM}$ are negative (positive) for the structures STO/AlO/YBCO and AlO/YBCO/STO (YBCO/STO/AlO, STO/YBCO/AlO, and YBCO/AlO/STO) for all incidence angles, as depicted with dashed line in Fig. 3(a) and dotted line in Fig. 3(b) (solid and dotted curves in Fig. 3(a) and solid line in Fig. 3(b)), respectively. For the structure AlO/STO/YBCO $\Delta X_{TM} < 0$ for $\theta < 64°$, then it changes sign and also becomes significant only at grazing incidence, as shown with dashed line in Fig. 3(b). These dependencies are in agreement with calculations of lateral shift for dielectric/metal interface reported earlier [31, 32, 36]. In contrast to those cases, in the systems with SC layer instead of pure metal, properties of the reflected waves are strongly dependent on temperature.

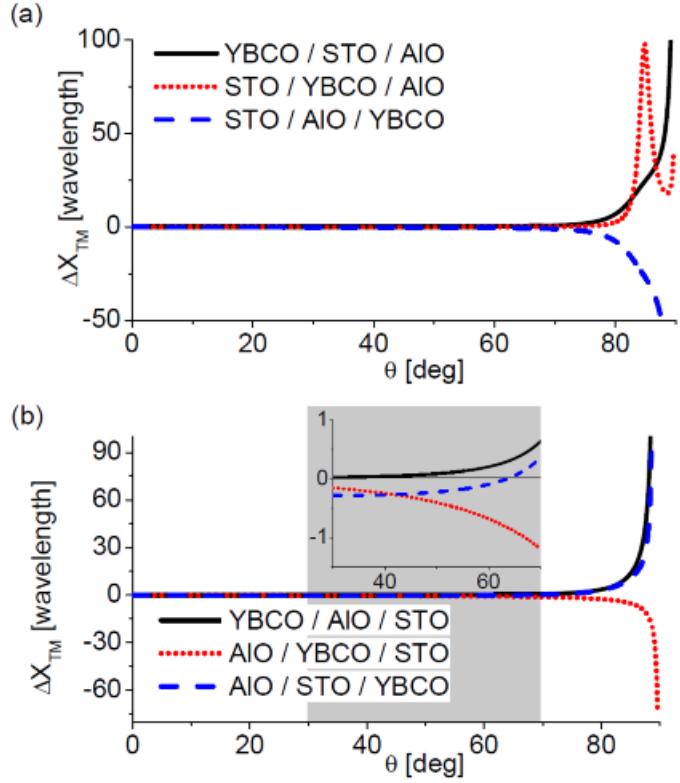

FIG. 3. Dimensionless lateral shifts $\Delta X_{TM}$ vs incidence angle $\theta$ for different positions of the SC film for TM-polarized electromagnetic wave incident in (a) STO, AlO and (b) AlO, STO dielectric layers (in order of light passing though the structure). The thickness of the SC film is $d_{YBCO}$ = 50 nm, the angular frequency $\omega$ = 100 rad×THz.

## B. Governing the lateral shift for the structure YBCO/STO/AlO by temperature and the electromagnetic wave parameters

We investigate dependence of the lateral shift of TE-wave on temperature, the thickness of the SC layer and the angular frequency and the incidence angle of the electromagnetic wave. We restrict our analysis to study the behavior of the $\Delta X_{TE}$ peak at pseudo-Brewster angle for the structure YBCO/STO/AlO.

In Fig. 4 we present the color map of the $\Delta X_{TE}$ peak's evolution with the incidence angle $\theta$ and temperature $T$. The color shows the values of the lateral shift. The thickness of the SC film is $d_{YBCO}$ = 50 nm, the angular frequency $\omega$ = 100 rad×THz. We investigate the temperature range from zero to the SC transition temperature of YBCO ($T_C$=90 K). The increase of temperature leads to decrease of $\Delta X_{TE}$ maximum due to growing absorbance in the SC layer. This tendency is in agreement with the results reported for the lateral shift in a lossy medium [37]. Also the peak's position slightly drifts with temperature growth towards larger $\theta$, because changing of the permittivity $\varepsilon_{yy}$ with temperature results in shift of the pseudo-Brewster angle. It should be noted that decreasing of $\Delta X_{TE}$ with the temperature growth in the vicinity of the pseudo-Brewster angle in system with a SC layer is opposite to behavior of the lateral shift in dielectric/metal composites in high ($T > 300$ K) temperatures [38]. As one can see from Fig. 4, the lateral shift is sufficient not only at liquid helium temperatures. At temperatures around liquid nitrogen (T = 77 K) the lateral shift is around 15 wavelengths, which is also can be detected experimentally. It should be noted that nowadays the precise optical experiments can be performed in ultralow temperatures. For example, recently in Ref. [39] it was reported about optical measurements of a nanoscale silicon optomechanical crystal cavity with mechanical resonances at subkelvin temperatures. Also in Ref. [40] the experimental equipment for light reflection measurements in infrared regime at liquid helium temperatures has been reported.

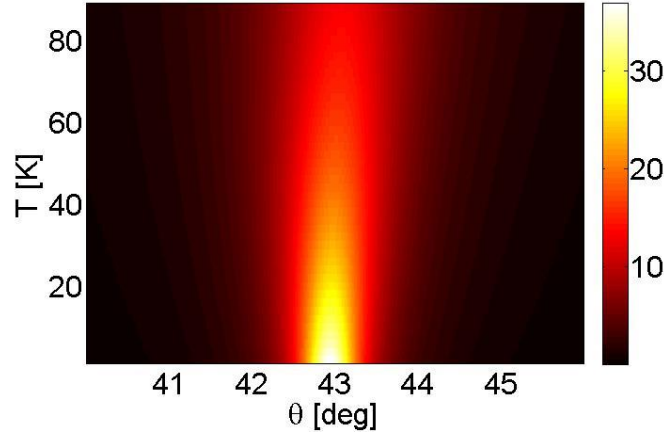

FIG. 4. Color map of the dimensionless lateral shift $\Delta X_{TE}$ evolution with the incidence angle $\theta$ and temperature $T$ for the structure YBCO/STO/AlO. The thickness of the SC film is $d_{YBCO}$ = 50 nm, the angular frequency $\omega$ = 100 rad×THz.

In Fig. 5 one can see the color map of the $\Delta X_{TE}$ peak's evolution with $\theta$ and the SC layer thickness $d_{YBCO}$ within the range (0 ÷ 70) nm. Here the temperature is $T$ = 5 K, and the angular frequency $\omega$ = 100 rad×THz. The peaks of the lateral shift broaden and drift towards smaller incidence angles with the increase of $d_{YBCO}$. Such a behavior is caused by the changes in the complex reflection coefficient $R_{TE}$ due to presence of the SC film. The introduction of the SC sublayer leads to the shift of the pseudo-Brewster angle towards smaller $\theta$ and to the increase of the reflection coefficients values at the pseudo-Brewster angle, providing the considerable changes in the phase difference $\psi$ at the reflection. The maximal lateral shift takes place for the pure dielectric bilayer STO/ALO at the limiting case of $d_{YBCO}$ = 0 (the absence of the SC layer), though this could be hardly observable as in this case reflection coefficient tends to zero at the Brewster angle. The increase of $d_{YBCO}$ leads to the absorption growth, thus $\Delta X_{TE}$ decreases, as one can see from Fig. 5.

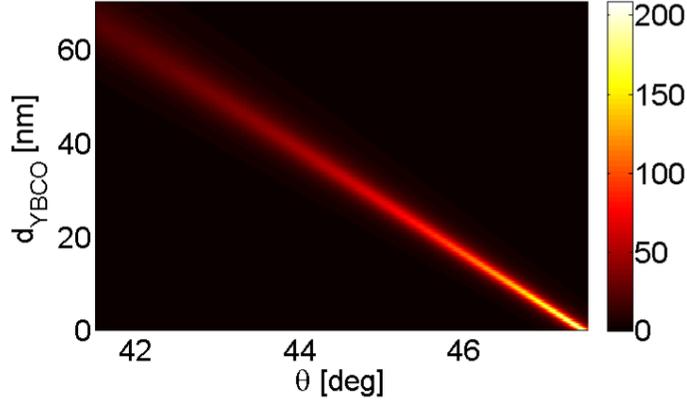

FIG. 5. Color map of the dimensionless lateral shift $\Delta X_{TE}$ as function of the incidence angle $\theta$ and SC layer thickness $d_{YBCO}$ for the structure YBCO/STO/AlO. The temperature is $T = 5$ K, the angular frequency $\omega = 100$ rad×THz.

In Fig. 6 the dependence of $\Delta X_{TE}$ on $\theta$ for different angular frequencies $\omega = 96, 100, 105,$ and 107 rad×THz is shown. The increase of $\omega$ results in the sharpening of the lateral shift peaks and growth of its maximal values and to the drift of the peaks towards the larger $\theta$. According to our estimations, at $\omega = 96$ rad×THz the maximal value of lateral shift corresponds to the incidence angle $\theta = 20°$ and its value does not exceed ten wavelength. With changing of the incident light frequency to $\omega = 100$ rad×THz the lateral shift peak shift to $\theta \approx 43°$, and the maximal lateral shift value already reaches several tens of wavelength. The further increase of the incident light frequency to 105 and 107 rad×THz makes the lateral shift maxima to drift to $\theta \approx 65°$ and $\theta \approx 78°$, respectively, which is accompanied by the drastic growth of $\Delta X_{TE}$ up to several hundred of wavelength (See Fig. 6). The obtained pronounced frequency dependence of the lateral shift originates from the strong frequency dispersion of the permittivity component $\varepsilon_{yy}$.

It should be noted that the dependencies of $\Delta X_{TE}$ on temperature, frequency and SC layer thickness in the structure YBCO/STO/AlO have similar behavior for other configurations, where the Goos-Hänchen shift has peak at pseudo-Brewster angle, namely, STO/YBCO/AlO and STO/AlO/YBCO.

As it is shown in Ref. [17] in the case of SC permittivity isotropic in *xy*-plane, the lateral shift of the light transmitted through a photonic crystal with SC constituents is relatively small in comparison with the case of SC permittivity anisotropic in *xy*-plane. For the three-layered systems under consideration, for the SC permittivity isotropic in the film plane, $\Delta X$ has no pronounced maxima and it increases at the grazing incidence only. The behavior of the lateral shift for both TE- and TM-polarized light is similar to presented in Fig. 3(b).

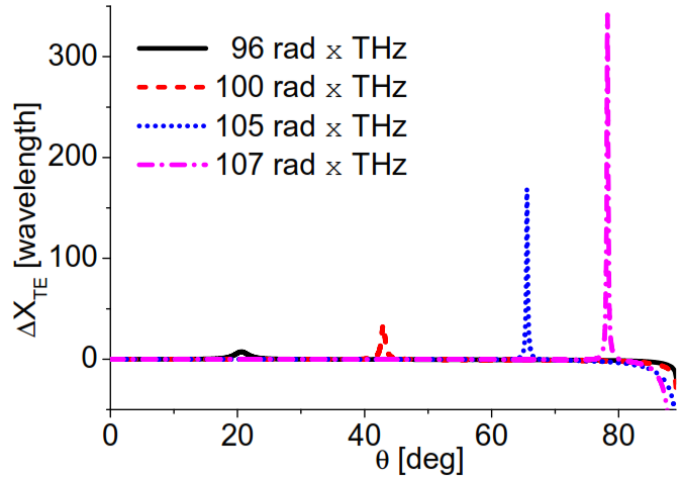

FIG. 6. Dimensionless lateral shifts $\Delta X_{TE}$ as functions of the incidence angle $\theta$ for different angular frequencies: 96, 100, 105, and 107 rad×THz (solid, dashed, dotted, and dash-dotted lines, respectively), for $T = 5$ K and $d_{YBCO} = 50$ nm.

## IV. CONCLUSIONS

In conclusion, we have investigated the lateral shift of the light, reflected from the complex superconducting–dielectric three-layer structures and dependence of the value of this shift on the mutual location of these layers. We have shown that the lateral shift is very sensitive to the anisotropy of the optical properties of $YBa_2Cu_3O_7$ film, as well as to the position of the SC layer in the three-layer system. We also showed that varying the temperature and choosing the position of the superconducting film it is possible to obtain both positive and negative large values of the lateral shift of TE-polarized reflected light at the incidence angles near the pseudo-Brewster angles, whereas for TM-polarized electromagnetic wave, the lateral shift is only considerable at the grazing incidence. As follows from aforementioned analysis, tuning such parameters as the temperature, incidence angle and the incident light frequency, one can govern the Goos-Hänchen effect. We believe that our results can be useful for investigation of lateral shift in complex structures containing superconducting constituents, particularly in dielectric photonic crystals with superconducting defect layers [41 – 47].

## ACKNOWLEDGMENTS

This research has received funding from the European Union's Horizon 2020 research and innovation program under the Marie Skłodowska-Curie grant agreement No 644348, project "MAGIC" (N.N.D., Yu.S.D., and I.L.L.), and COST Action MP1403 "Nanoscale Quantum Optics" (N.N.D., Yu.S.D., and I.L.L.) and also is supported by the grants from Ministry of Education and Science of Russian Federation: State Assignment and Project No. GK RFMEFI57414X0057 (N.N.D. and Yu.S.D.).